\documentclass[epj]{webofc}
\usepackage[varg]{txfonts}   
\woctitle{ICNFP 2017}
\newcommand{\beq}{\begin{equation}}
\newcommand{\eeq}{\end{equation}}
\newcommand{\bea}{\begin{array}}
\newcommand{\eea}{\end{array}}
\newcommand{\beqa}{\begin{eqnarray}}
\newcommand{\eeqa}{\end{eqnarray}}

\begin{document}
\selectlanguage{english}
\title{Chiral fluids: a few theoretical issues }

%
%

\author{V.~I, Zakharov\inst{1,2}\fnsep\thanks{\email{vzakharov@itep.ru}} \and
        O.~V. Teryaev\inst{3,4}  
}


\institute{ NRC Kurchatov Institute - ITEP, 117218 Moscow, Russia,
\and
           School of Biomedicine, Far Eastern Federal University, 690950 Vladivostok, Russia,
\and
           JINR           
\and
            Moscow  Institute of Engineering
}

\abstract{%
 We review briefly a few topics concerning physics of fluids whose constituents
are massless fermions interacting in chiral invariant way. Macroscopic
 manifestations of the chral anomaly is one of central issues.
Another topic is ultraviolet vs infrared sensistivity of chiral magnetic and vortical effects.
To clarify dynamical issues involved we rely mostly on a (well-known) 
toy model of pionic superfluidity.  
}
\maketitle

\section{Introduction}
\label{sec:introduction}

By chiral fluids one understands fluids whose constituents are massless
fermions. In presence of external 
electric and magnetic fields there arises the chiral anomaly, as usual.
Manifestations of the chiral anomaly in the standard set up of field theory
have been exhaustively studied. If instead of the vacuum-to vacuum matrix elements
one turns to the matrix elements over medium
the expectation would be--we guess--that the effect of the anomaly
is damped down. Indeed, the anomaly is a quantum effect and is sensitive to phases
of amplitudes. Interaction with medium would lead to the loss of
coherence and wash out the effect of the anomaly. Superfluidity is an exceptional 
and well understood case when quantum effects survive macroscopically.

 On the background of such intuitive picture the results of an explicit 
evaluation of the effect of the anomaly in the
hydrodynamic approximation \cite{Sons:2009}
turned unexpected. Namely, 
It was found that the effect of anomaly survives and 
leads to macroscopic, so called chiral effects. We mention here
two such effects, or  novel contributions to the vector and axial-vector currents
 $j_{\alpha}$ and $j_{\alpha}^5$, respectively:
\begin {eqnarray}\label{chiral}\nonumber
j_{\alpha}~=~n_Vu_{\alpha}~+~C_{anom}e^2\mu_AB_{\alpha}~,\\
j^5_{\alpha}~=~n_Au_{\alpha}+~C_{anom}(\mu_V^2+\mu_A^2)\omega_{\alpha}~,
\end{eqnarray}
where $u_{\alpha}$ is the 4-velocity of an element of the fluid,
 $n_V, n_A$ are the sum and difference of the densities of the right- and
left-handed constituents, $\mu_V, \mu_A$  are the corresponding chemical potentials, 
$B_{\alpha}$ is the electromagnetic field in the medium, $B_{\alpha}=
(1/2)\epsilon_{\alpha\beta\gamma\delta}u^{\beta}F^{\gamma\delta}$ 
and $\omega_{\alpha}$ is the vorticity, $\omega_{\alpha}=
(1/2)\epsilon_{\alpha\beta\gamma\delta}
u^{\beta}\partial^{\gamma}u^{\delta}$, $e$ is the electromagnetic coupling.
Finally, $C_{anom}$ is a constant which appears in front of the anomaly
and its presence signifies the fact that the corresponding term 
would be absent in absence of the anomaly.  The value of the constant $C_{anoma}$
depends on the number of massless constituents and 
$C_{anom}=1/(2\pi^2)$ in case of a single Dirac particle of charge $e$. All the chiral effects
are proportional to $C_{anom}$.

This amusing conclusion that the effect of the anomaly survives in the hydrodynamic limit
is not the only feature which triggerred a lot of interest in  the
chiral effect. The first years of the development were summarized in a volume
of reviews \cite{volume:2013}. Here, we will pick up only two points.

Turn first to chiral magnetic effect, see the term proportional to $C_{anom}$ 
in the first line of 
the expression (\ref{chiral}). At least at its face value, the
chiral magnetic effect is a flow of electric current along an
external magnetic field. Then one can argue \cite{Kharzeev:2011}
that this current is dissipation-free (in a CP-invariant theory).
The proof is straightforward and is based on the observation that 
the chiral magnetic current and the standard electric current have opposite
parities under reflection of time.  Thus, the chiral magnetic effect seems to echo superfluidity.
The puzzle is that there is no microscopic mechanism known which would  
result in dissipation-free transport in case of the chiral magnetic current.

The other observation in point is that the currents (\ref{chiral}) unify helical motions
of macroscopic and microscopic degrees of freedom. Indeed, concentrate on
the zeroth component of the axial current (\ref{chiral})
\begin{equation}\label{charge}
j_0^5~=~ (n_R-n_L)u_0 +C_{anoma}(\mu^2+\mu_5^2)\vec{\Omega}\cdot \vec{v}~,
\end{equation}
The first term in the right side is the difference
of densities of right- and left-handed constituents.
While the other term is non-vanishing for a macroscopic helical motion.
Furthermore, one can argue 
\cite{Avdoshkin:2016,Yamamoto:2016} that unification of these terms eventually results in
possiblity of transition between the two terms (so that the total charge is conserved).
There arises a similar mixing between helicity of macroscopic magnetic fields and
difference in number of right- and left-handed photons. In fact, it is the
instability induced by the mixing of the two electromagnetic terms which was 
studied in most detail \cite{Redlich:1984,Akamatsu:2013,Khaidukov:2013}.

Again, the mechanism of the transition between the microscopic and macroscopic motions
is difficult to visualize. Indeed, chirality of the massless constituents is defined 
without any reference to externally determined directions. A particle remains, say, left-handed
while moving in various directions. On the other hand, the macroscopic helical motion
is triggerred by external forces. The macroscopic motion can be imitated by
microscopic degres of freedom only if the massless particles 
start moving predominantly in
a particular direction. To our knowledge, there is no
mechanism to produce such an alignment
of the momenta of the constituents.

In conclusion of this brief introduction let us emphasize that the literature
on the chiral fluids  is huge
and in no way is referred to here systematically. We cite papers mostly in cases
when we reproduce directly argumentation of these papers. Otherwise, the reader
is referred to the corresponding original papers as well as review papers for 
more comprehensive lists of references.

\section{Extra conservation laws in case of ideal fluids}
Equations (\ref{chiral}) have been reproduced many times, by different techniques. 
The standard derivation of (\ref{chiral}) starts with external elctromagnetic fields present.
In the final answer one is free to put fileds to zero. Note, however, that 
the expression (\ref{chiral}) for the axial current contains the chiral vortical effect
also in absence of external fields. Therefore,
we come to a paradoxical conclusion that   appearance of $C_{anom}$ in front
of a particular term in the current 
is not actually a proof of its relation to the anomaly!

An alternative derivation of the anomalous terms which is valid also in absence of 
the external fields was suggested in Ref. \cite{Sadofyev:2011} . The point is
that in thermodynamics one introduces effective Hamiltonian $\hat{H}_{eff}$
which is related  to the original Hamiltonian of the system $\hat{H}_0$ as:
\begin{equation}\label{extension}
\hat{H}_{eff}~=~\hat{H}_0~-~\mu\cdot \hat{Q}~~,
\end{equation}
where the chemical potential $\mu$ is conjugated to the charge $Q$.
In the hydrodynamic set up, the extension (\ref{extension}) corresponds to
introduction of effective Lagrangian:
\begin{equation}\label{modification}
\delta L_{eff}~=~\mu u_{\alpha}j^{\alpha}~~,
\end{equation}
where$\int d^3xj^0\equiv Q$.

Imagine now that we start calculating matrix elements of the currents in the
hydrodynamic approximation beginning with the fundamental Lagrangian
modified according to (\ref{modification}). Applying then the Adler-Bardeen theorem
we would uniquely fix an anomalous contribution
which can be generated from the standard anomaly by the
substitution \cite{Zakharov:2016}
\begin{equation}\label{substitution1}
eA_{\alpha}~\to~eA_{\alpha}+\mu u_{\alpha} ~.
\end{equation}
One can readily check that in this way one reproduces the standard chiral effects.
  
This completes the derivation of the chiral vortical effect in absence of external fields.
Solving one problem, however, brings us to confront another problem.
Namely, if one evaluates directly Feynman graphs the
bare triangle graph is singled out by nonconservation of
the axial current. In case of the chiral vortical effect:
\begin{equation}\label{wrong}
\big(\partial^{\alpha}j^5_{\alpha}\big)_{vortical}~\sim~\vec{\Omega}\cdot \vec{a}~~,
\end{equation}
$\Omega_i=\frac{1}{2}\epsilon_{ijk}\partial^jv^k$, $a_i~=dv_i/dt$ and we 
use non-relativistic approximation for simplicity (c.f. \cite{Prokhorov:2017atp}).
However, it is well known that introduction of a chemical potential or temperature
does not affect the chiral anomaly. Indeed, the anomaly can be derived
entirely in terms of UV regularization which is not sensitive to introduction of a medium.

A way to resolve this paradox is to realize that the r.h.s. of Eq. (\ref{wrong})
does not vanish on the equations of motion which follow from
the fundamental Lagrangian. But the accelaration $\vec{a}~=~0$ for the fluid in
equilibrium, see, in particular \cite{Moffat:1969,Bekenstein:1987}. 
Thus, the matrix element (\ref{wrong}) vanishes in the equilibrium,
or on the equations of motion for the medium. Note that in the limit
of ideal fluid the velocities $\vec{v}$ do not vanish, generally speaking, and the fluid
can have a nontrivial, conserved helical charge, see Eq.  (\ref{charge}). Introduction
of dissipation would generically lead finally to $\vec{v} = 0$. Thus, the helical charge
is not conserved for viscous fluid. This observation is not in contradiction 
with our derivation of the chiral vortical effect in terms of the Feynman graphs.
Indeed, unitary field theory cannot incorporate dissipation.

\section{Ideal fluid in external gravitational field}
\subsection{Conservation laws}
In this section we consider motion of chiral fluid in an external gravtational field.
One of purposes of introduction of the gravitational field is 
to address theory of thermal chiral effects.
Indeed, according to Luttinger \cite{Luttinger:1964} external gravitational field imitates the effect of a
nonvanishing gradient of temperature. Therefore, studying chiral fluids in
external gravitational filed can be useful to appreciate chiral thermal effects,
see \cite{Basar:2013} and references therein.  

Let us concentrate on conservation laws for the axial current,
see (\ref{chiral}). As we argued in the preceding section,
in case of ideal fluid (and absence of external fields)
there are actually two separately conserved currents:
\begin{eqnarray}\label{twolines}\nonumber
\partial_{\alpha} (n_Au^{\alpha})~=~0~, \\
\epsilon^{\alpha\beta\gamma\delta}\partial_{\alpha}
(\mu u_{\beta})\partial_{\gamma}(\mu u_{\delta})~=~0 .
\end{eqnarray}
The first line here is the conservation of the standard hydrodynamic current
which is granted since we do not include external electromagnetic fields.
The second line is then consequence of (\ref{chiral}). The physical meaning
of the latter conservation law is the vanishing acceleration, $\vec{a}=0$,
see for the discussion above.

As the next step, switch on an external gravitational field. 
The ordinary derivatives $\partial_{\alpha}$ are replaced 
then by  covariant ones, $D_{\alpha}$ and for any current $j_{\alpha}$
conserved in the flat-space limit one gets:
\begin{equation}
\partial_{\alpha}j^{\alpha}~=~0~\to~D_{\alpha}\mathcal{J}^{\alpha}~=~0~,
\end{equation}
where the current $\mathcal{J}_{\alpha}$ transforms as a vector 
under general coordinate transformations.
In the flat-space limit the current $\mathcal{J}_{\alpha}$ coincides
with $j_{\alpha}$. In curved space, there are deviations of $\mathcal{J}_{\alpha}$ from
$j_{\alpha}$ which are fixed by the transformation properties of $\mathcal{J}_{\alpha}$
under coordinate change.

In presence of gravitational field the conservation law now looks as: 
\begin{equation}\label{divergence}
\mathcal{J}^{\mu}_{;\mu}~=~\frac{\partial \mathcal{J}^{\mu}}{\partial x^{\mu}}+
\Gamma^{\mu}_{\mu\rho}\mathcal{J}^{\rho}~=~0,
\end{equation}
where $\mathcal{J}^{\mu}_{;\mu}\equiv D_{\mu}\mathcal{J}^{\mu}$ and 
$\Gamma^{\mu}_{\nu\rho}$ are Christoffel symbols.
We will be interested in the lowest-order effect of gravity
and assume that deviations from the Minkowskian space are weak:
$g_{\mu\nu}=\delta_{\mu\nu}+h_{\mu\nu},~~|h_{\mu\nu}| \ll  1$.
 In case of a weak field the general expression (\ref{divergence}) becomes:
\begin{equation}\label{weak}
\partial_{\mu}\mathcal{J}^{\mu}~\approx~-(1/2)\mathcal{J}^{\mu}\partial_{\mu}h~~,
\end{equation}
 where $h~\equiv~h_{00}-\Sigma_ih_{ii}$.
The $h_{00}$ component, as usual,  is fixed by the equivalence principle:
\begin{equation}\label{h00}
g_{00}-1~\approx~2\phi_{gr}~\approx~-2\vec{a}\cdot \vec{x}~.
\end{equation}
where $\vec{a}$ is the acceleration. 
 
Applying  these general relations to our case (\ref{twolines})
we get to first order in gravtational field:
\begin{eqnarray}\label{quasianomaly}
\partial_{\alpha}(n_Au^{\alpha})~=~0,\\\nonumber
\partial_{\alpha}\epsilon^{\alpha\beta\gamma\delta}(\mu u^{\beta})\partial^{\gamma}
(\mu u^{\delta})~=~(const)\mu^2 \vec{\Omega}\cdot\vec{a}
\end{eqnarray}
Note absence of  corrections to the standard hydrodynamic current in the ideal-fluid limit. It
is a consequence of the relation $u^{\alpha}a_{\alpha}=0$ where $a_{\alpha}=(d/ds)u_{\alpha}$.
Indeed, if the fluid is at rest, then $(d/dt)(nu_0)~=~0$ according to this relation.
For ideal fluid, generally speaking, there is no universal rest frame where one could apply this relation.
However, for the ideal fluid the symmetry group includes diffeomorphism
and charge conservation is obeyed separately within each comoving fluid
element
(for an enlightening  
discussion of conservation laws for ideal fluids see \cite{Dubovsky:2012}).
 
 \subsection{Thermal vortical effect} 

As is noted in Ref. \cite{Basar:2013} consideration of fluid conservation laws in
external gravitational field allows to derive novel thermal chiral effects.
One simply replaces the acceleration 
(in the appropriate units system) induced by the gravitational field
by 
\begin{equation}\label{acceleration}
\vec{a}_{gr}~=~-\frac{\vec{\nabla}T}{T}~~,
\end{equation}
as first proposed by Luttinger \cite{Luttinger:1964}.

Following this suggestion we get for divergence of the fluid-helicity current:
\begin{equation}\label{thermal}
\partial_{\alpha}\epsilon^{\alpha\beta\gamma\delta}(\mu u^{\beta})\partial^{\gamma}
(\mu u^{\delta})~=~-(const)\frac{\mu^2}{T} \vec{\Omega}\cdot\vec{\nabla}T
\end{equation}
This equation is very similar to that obtained in \cite{Basar:2013}
for the same type of helical motion, with acceleration $\vec{a}$ being parallel to
the angular velocity $\vec{\Omega}$, see Eq. (\ref{quasianomaly}). There is
an important difference as well, however. Namely, in Ref. \cite{Basar:2013}
the r.h.s. of Eq. (\ref{thermal}) stands for the divergence of the total
axial current, see (\ref{chiral}). Let us remind the reader that
the two terms in Eq. (\ref{chiral}) for the axial current
look diffierent mcroscopically. Namely, the standard hydrodynamic current refers to
the flow of elementary chiral coinstituents while the fluid-helicity current
 describes macroscopic helical motion. Thus, the result of Ref. \cite{Basar:2013} 
allows for interpretation that
it is the elementary constituents of certain chirality flowing 
along the vector $\vec{\nabla}T$ that are responsible for 
a nonvanishing divergence of  $j_{\alpha}^5$. 
While in our case, see Eq. (\ref{thermal}), there is no hint on 
mixing of the microcopic and macroscopic
degrees of freedom, for further details and references see \cite{Minkin:2017}.

It is worth mentioning that the original proposal of Luttinger \cite{Luttinger:1964}
referred to hydrodynamic motion in resistive media. In other words, 
there is flow of fluid along the gradient of the gravitational
potential $\vec{\nabla}\phi_{gr}$ but with a constant velocity,
without acceleration,  $\vec{a}_{fluid}=0$.  
On the other hand, substitution (\ref{acceleration}) proposed
in Ref. \cite{Basar:2013} refers to the dissipation-free motion, and is valid in
 high-frequency limit. Validity of this approximation 
is to be justified in each particular case.

\subsection{Emergent gravity}

Turn back to consideration of motion in flat space. For ideal fluid
in equilibrium, the fluid helicity current, see  Eq. (\ref{quasianomaly})
is conserved. However, for viscous fluids there is no such conservation law
\cite{Moffat:1969,Bekenstein:1987}. Indeed, for a viscous fluid the equilibrium
configuration in most cases is nothing else but the whole of the fluid being at rest
in a particular frame so that the helical charge vanishes. 

It is speculated in \cite{Zakharov:2016} that at
short times the effect of non-conservation of 
helicity can be described in terms of effective gravity. In more detail, let us start
with motion of ideal fluid and then switch on non-vanishing viscosity. At short times, there
appear acceleration which is proportional to the viscosity and we get
a non-vanishing r.h.s. of Eq, (\ref{quasianomaly}). This acceleration
is fixed in terms of the initial conditions of the motion of the ideal fluid.
Furthermore, one could introduce effective gravitational field
which reproduces this acceleration. In 
other words, the emergent gravitational field is fixed in terms 
of the motion of dissipative fluid. In this approximation,
the conservation of the fluid helicity holds in terms of the respective covariant derivatives,
see discussion above.

This qualitative picture does not contradict much more systematic 
general studies of theory of dissipative fluids.
Ref. \cite{Crossley:2015}, e.g.,  introduces both an effective metric tensor, defined in terms 
of the motion of the fluid itself,
and supersymmetry, responsible for the dissipation.

 \section{Ultraviolet vs infrared sensitivity of the chiral effects.}
\subsection{Pion superfluidity}

In this section we review a simple toy model of chiral effects.
One starts with pionic medium in the presence of chemical potentials,
$\mu_{3,V},\mu_{3,A}$ violating isotopic symmetry. The corresponding 
Lagrangian in terms of the quark fields is given by:
\begin{equation}\label{model}
\delta L~=~\mu_{3,V}\bar{q}\gamma_0(\tau_3/2)q~+~
\mu_{3,A}\bar{q}\gamma_0\gamma_5(\tau_3/2)q ~,
\end{equation}
where $q$ are isotopic doublets, quark fields, 
$\gamma_0,\gamma_5$ are Dirac matrices,$(\tau^3/2)$
is the generator of rotations around the third axis in the isotopic-spin space. 
It is well known that in the low-energy domain the model (\ref{model}) exhibits pionic
superfluidity. Although the derivation of this amusing result is absolutely straightforward,
because of the space consideration we only sketch it. Details and further references can be found in
\cite{Son:2001,Aharony:2007}. Furthermore,
 switching on external magnetic fields and/or rotation one can derive
the chiral effects which is our central point in this section.
In presentation we follow \cite{Teryaev:2017,Avdoshkin:2017}
 where further details and references can be found.

The proof of superfluidity is given in terms of effective Lagrangians
describing interaction of light degrees of freedom, or pions.
Goldstone fields, or pions represent phases of a unitary matrix $U$:
\begin{equation}\label{umatrix}
U~=~\exp (i{\tau^a\pi^a}/{f_{\pi}})~,
\end{equation} 
where $f_{\pi}$ is a constant, $\tau^a (a=1,2,3)$ are Pauli matrices,
$\pi^a$ are pion fields.
The leading term in the effective Lagrangian has two derivatives
and corresponds to the kinetic energy of the pions. 
To incorporate the effect of the chemical potentials one
replaces ordinary derivatives by the covariant ones \cite{Son:2001}:
\begin{equation}\label{lchirallagrangian}
L_{chiral}=~\frac{f_{\pi}^2}{4}\Big(D_{\mu}UD^{\mu}U^{\dagger}\Big)~,
\end{equation}
where the covariant derivatives are defined as:
\begin{eqnarray}
D_{\mu}U~=~\partial_{\mu}U
~-~i\delta_{\mu 0}\big(\hat{\mu}_{3,L}U-U\hat{\mu}_{3,R}\big)~~,
\end{eqnarray}
and $\hat{\mu}_{3,L}, \hat{\mu}_{3,R}~
\equiv~\mu_{3,L}\sigma_3, \mu_{3,R}\sigma_3$.   The chemical potentials $\mu_{3,L},\mu_{3,R}$
are related to $\mu_{V,3},\mu_{A,3}$ introduced in Eq. (\ref{model}) in an obvious way.

In fact, the cases $(\mu_{3,V}\neq 0, \mu_{3,A}=0)$ and $(\mu_{3,V}=0, \mu_{3,A}\neq 0)$ 
in the limit of exact chiral symmetry
can be reduced
to each other by a change in notations. The point is explained in detail in Ref. \cite{Aharony:2007},
 We  will consider
$\mu_{3,A}\neq 0, \mu_{3,V}=0$.
Then the
potential energy  $V_{chiral}$ corresponding to the Lagrangian
(\ref{lchirallagrangian}) is given by :
\begin{equation}\label{potential}
V_{chiral}~=~-\frac{f_{\pi}^2}{8}\mu_{3,A}^2\big(Tr (U\sigma_3U^{\dagger}\sigma_3)
+2Tr I~\big)~,
\end{equation}
The minimum of the potential energy
is reached on matrices $U_{min}$ where
\begin{equation}
U_{min}~=~I \cos\phi~+~\sigma_3 \sin\phi  ~.
\end{equation}
The independence of the energy (\ref{potential}) on 
the angle $\phi$ 
implies presence of a Goldstone boson.
 The superfluid solution is associated with the phase growing with time:
\begin{equation}\label{cartan}
U_{solution}~=~\exp \big(i\sigma_3\mu_{3,A}\cdot t\big)~~,
\end{equation}
In other words,
the field $\pi_0$, defined in (\ref{umatrix}) looks as, 
\begin{equation}\label{fieldpi}
\frac{\pi_0}{f_{\pi}}=\mu\cdot t+\varphi(x_i),
\end{equation}
where the function $\varphi(x_i)$ satisfies the equation $\Delta\varphi(x_i)=0$.
The 3d massless field $\varphi$ signifies superfluidity.
 
Differentiating the potential energy with respect to $\mu_{3,A}$
one gets the density  of pions in the ground state:
\begin{equation}\label{density}
n_5~=~f_{\pi}^2\mu_{3,5}~~.
\end{equation}
A similar equation holds if one starts with $\mu_{3,V}\neq 0, \mu_{3,A}=0$.

As we argue in the next subsection, knowing Eq. (\ref{density}) allows for 
a quick derivation of the chiral magnetic effect.

\subsection{Chiral effects in superfluid as radiative corrections}

In this subsection we review theory of the chiral magnetic effect treated as a radiative 
correction to the relation (\ref{density}). One first rewrites (\ref{density}) as
an expression for the hydrodynamic axial current:
\begin{equation}\label{hydro}
j_{\alpha}^5~=~f_{\pi}\partial_{\alpha}\pi^0~=~n_5u_{\alpha}~,
\end{equation}
where $n_5$ is the density of $\pi^0$-mesons. And then one accounts
 for the interaction of $\pi^0$ with external electromagnetic fields,
following the standard rules of quantum field theory. Similar approach was
exploited in, e.g., 
\cite{Son:2008,Lublinsky:2010,Kalaydzhyan:2014}. We follow closely
\cite{Teryaev:2017,Avdoshkin:2017} and argue that in the hydrodynamic approximation
evaluation of the chiral magnetic effect is factorized into calculation
of the density (\ref{density}) and of the amplitude of interaction of $\pi^0$-meson with
external electromagnetic fields. 

Decay of $\pi^0$ into two photons is governed by an effective vertex:
\begin{equation}\label{pi0}
\delta L~=~f_{\pi^0\to 2\gamma}\pi^0\epsilon^{\alpha\beta\gamma\delta}
\partial_{\alpha}A_{\beta}\partial_{\gamma}A_{\delta}~,
\end{equation}
where $f_{\pi^0\to 2\gamma}$ is a constant,
$\pi^0$ is the field of $\pi^0$-meson. Note that Eq. (\ref{pi0})
incorporates in fact the chiral anomaly. Indeed, if there were no anomaly
the vertex (\ref{pi0}) would have vanished  in the limit
of vanishing 4-momentum of the pion and $f_{\pi^0\to 2\gamma}$ 
could not be treated as a constant. By varying (\ref{pi0}) with respect to the potential $A_{\alpha}$
one can generate an effective electromagnetic current $j^{el}_{\alpha}$. Moreover,
it is useful first to integrate (\ref{pi0}) by parts. As a result,
\begin{equation}\label{elcurrent}
j^{el}_{\alpha}~=~f_{\pi^0\to 2\gamma}\epsilon_{\alpha\beta\gamma\delta}
(\partial^{\beta}\pi^0)F^{\gamma\delta}~,
\end{equation}
where $F^{\gamma\delta}$ is the electromagnetic field strength tensor. Note 
that the current (\ref{elcurrent}) was introduced first in Refs. \cite{Goldstone:1981,Callan:1985}.

To make contact with hydrodynamics (of superfluid) we need to only substitute
(\ref{hydro}) to (\ref{chiral}). It is readily seen that we do reproduce the
chiral magnetic effect since in case we consider the constant $C_{anom}$
entering Eq (\ref{chiral}) is given by
\begin{equation}
C_{anom}(4\pi \alpha_{el})~=~f_{\pi}\cdot f_{\pi^0\to 2\gamma}~~.
\end{equation}
And this equation completes the derivation of the chiral magnetic moment in
the  model considered.

Now, that we have an explicit calculation of a chiral effect we can come back to questions
mentioned in the Introduction. First, the mystery of dissipation-free transport is
resolved now through the observation that the chiral magnetic effect 
corresponds to a polynomial in the effective action. Thus, there is no actual ``transport''
behind this effect. (Note, however, that the displacement current that we have in our model
can be transformed into ordinary current on a subspace of lower dimension, see
in particular
\cite{Callan:1985}.)

Another application of the technique outlined above is the evaluation of dependence of
the chiral separation effect on pion mass. The chiral separation effect is associated with
$\mu_{3,V} \neq 0$. Non-vanishing pion mass can consistently be incorporated in this case
and $m_{\pi}\neq 0$ modifies in a well-defined way the density (\ref{density}) \cite{Son:2001}.
Basing on considerations outlined in the current subsection one immediately concludes
that the density of the pions in the ground state is the only source of the ``fast''
dependence of the chiral separation effect on the infrared-sensitive physics.
For further
details see \cite{Avdoshkin:2017}. 

\section{Chiral vortical effect}
\subsection{Duality between chiral vortical effect and spin of vortices}

There is another aspect of duality which is made manifested 
within field-theoretic approch.
Consider next the chiral vortical current, see second line of Eq. (\ref{chiral}).
At first sight it looks as a next term in the derivative expansion. 
Thus, one could expect that
specific hydrodynamic excitations, like sound waves, are responsible
for the dynamics of the chiral vortical effect. Within the model we consider, however,
the chiral votical effect can also be derived as a radiative correction to the
supercurrent associated with short distances
\cite{Teryaev:2017}.

The reason is that the chiral vortical effect is saturated by the
contribution of cores of vortices, not by long-wave hydrodynamic excitations,
as we explain now in more detail. The starting point is the 
well known obsevation that
rotation cannot be transferred directly to a superfluid. 
Indeed, Eq. (\ref{hydro}) demonstrates that the velocity of the supecurrent
$\vec{v}~\sim~\vec{\nabla}\varphi$ and $\vec{\nabla}\times \vec{v}~\equiv~0$.
is $\vec{j}^5~\sim \vec{\nabla}\varphi$ and $\vec{\nabla}\times\vec{j}^5~=~0$
  However, rotation is still possible if one allows for singularities, or vortices.
 Near the singularity,
\begin{equation}\label{vicinity}
\frac{\pi^0}{f_{\pi}}~=~\mu_5\cdot t+\kappa \theta~,
\end{equation}
where $\theta$ is the polar angle and $\kappa$ is integer.  As a result,
\begin{equation}
(\partial_x\partial_y-\partial_y\partial_x)\theta~=~2\pi\delta(x,y).
\end{equation}
The condition that $\kappa$ is integer follows from the quantum nature of superfluidity.
Namely, the phase
of the wave function is to be single-valued  and, therefore,
\begin{equation}\label{quantization}
\mu_5\oint v_idx^i~=~2\pi \mathcal{\kappa}~, ~~\mathcal{\kappa}=
1,2...
\end{equation}
In applications, $\kappa=1$ in most cases.

Note that
in the hydrodynamic approximation
 size of the 
core of the vortex shrinks to zero. Thus,  it is commonly said that
(\ref{quantization}) determines spin of vortices. The spin of vortices defined in this
way is proportional to the Planck constant, as it should be.
Spin of the vortex is to be distinguished from the
total angular momentum associated with a particular vortex. The latter can be estimated as
\begin{equation}\label{angular}
\mathcal{L}~=~\int_0^{d_{vortex}}dr \cdot (mass)\cdot r\cdot v_{\phi}~\sim~d_{vortex}^2
\end{equation}
where $d_{vortex}$  stands for  the distance to the next vortex.  
This angular momentum  is not quantized.

The  total number  of vortices, $n_{vortex}$ is determined by the condition
that the total angular momentum of a bucket with supefluid rotated with angular velocity
$\vec{\omega}$ is (approximately) the same as of a solid body of the same shape as the bucket,
see, e.g., \cite{Landau:1959}. As a result, see, e.g., 
 \cite{Kirilin:2012} and references therein:
\begin{equation}\label{number}
n_{vortex}~=~\frac{\mu_5}{\pi}\int_Ad^2x|\omega_z|~, 
\end{equation} 
Averaging over vortices locally allows then to introduce a
quasi-continuum picture.  
 
So far we used hydrodynamics
to identify the chiral vortical effect with spin of vortices. 
Field theoretically, we can evaluate the same effect as a radiative correction
to the superfluidity itself. The logic is the same as used above in case of the
chiral magnetic effect. In this way, one gets \cite{Teryaev:2017}: 
\begin{equation}\label{extraterm}
\delta j^5_{\alpha}~=~\frac{1}{4\pi^2f_{\pi}^2}
\epsilon_{\alpha\beta\gamma\delta}
(\partial^{\beta}\pi^0)(\partial^{\gamma}\pi^0\partial^{\delta}\pi^0)~.
\end{equation}
Moreover, the field $\pi^0$ in the vicinity of the axis of a vortex 
is given by Eq. (\ref{vicinity}). Substituting (\ref{vicinity}) into 
(\ref{extraterm}) we get a field theoretic expression for the chiral vortical effect
which fits perfectly well the hydrodynamic derivation of it, for further
details and references see \cite{Teryaev:2017}.

\subsection{ Chiral anomaly as ``anomalous spin'' relation}

For normalization, consider first heavy, i.e. non-relativistic, fermions
with spin $S=1/2$ at temperature $T=0$. In the rotating frame spins of the particles
are aligned with the vector of angular velocity $\vec{\Omega}$. Density of the spin is then:
\begin{equation}
<\vec{S}>~=~\rho \frac{1}{2}\vec{n}_{\Omega}~,
\end{equation}
where $\vec{n}_{\Omega}$ is the unit vector in the direction of $\vec{\Omega}$,
$\rho$ is the density of fermions, and the Planck constant $\hbar$ is put to unit.

 The chiral vortical effect was evaluated first by Vilenkin
\cite{Vilenkin:1979}. In case of
massless fermions:
\begin{equation}\label{interprete}
\vec{j}_5~=\frac{\mu^2}{2\pi^2}\vec{\Omega}
\end{equation}

Exploring this duality between (totally antisymmetric) fermion spin density and axial current  
one can interpret (\ref{interprete}), at least formally, by saying that
effectively the particles now look like fermions at rest with an
induced spin 
\begin{equation}\label{estimate}
\vec{S}~=~\vec{\Omega}\frac{1}{|\mu|2\pi^2}~~,
\end{equation}
so that integration over the Fermi sphere produces the spin density ({\ref{interprete}).
Note that 
the r.h.s. is in fact proportional to the Planck constant
since it is associated with a loop graph
(which is similar to the standard triangle anomalous graph, with
the substitution (\ref{substitution1}) for the vertices).

Note  that the relation between axial current and spin density tensor  $S^{\mu, \nu  \lambda}$ for Dirac field is due to the fact that it is totally antisymmetric  and dual to the axial current. 
\begin{equation}
  S^{\mu, \nu  \lambda} =\frac{1}{2} \epsilon^{\mu \nu  \lambda \alpha} J_{5 \alpha},
\end{equation}
resulting in  the relation for the space components of axial current and spin density
\begin{equation}
\vec s = \vec{j}_5.
\end{equation}
The expression for spin per particle $\vec S \sim \vec \Omega / \mu$ compartible with (\ref{estimate})
may be obtained by assuming that particle occupy a Fermi sphere of radius $\mu$ with the density fixed by uncertainty relation. The growth of baryon polarization with chemical potential \cite{Sorin:2016smp}
compatible with experimental data may signal that this density is in fact not reached.  

Relation of the
anomaly to a kind of chiral-invariant magnetic or spin moment
has been studied in a number of papers, see, e.g., \cite{Son:2013}
and references therein.
Concentrate for a moment on case of magnetic moment of a charged
fermion. If one starts from a massive fermion, its magnetic moment
is inverse proportional to its mass. However, the use of such a description assumes
that the momentum transfer from magnetic field, $|\vec{q}|$ is much smaller
than mass,  $|\vec{q}|\ll m$. If one sends the mass to zero first,
to approach the chiral limit, then the magnetic moment does not survive since
the magnetic-type
coupling violates chirality conservation. However, if one includes the effect of the
anomaly, or Berry phase \cite{Son:2013}	
then there arises coupling of the massless fermion
to magnetic field, as if the fermion 
has chiral-invariant magnetic moment directed along the momentum
and inverse proportional to its energy. 

We expect that in a rotating coordinate system the total angular momentum
$\vec{\mathcal{J}}$ is aligned with the vector of angular velocity $\vec{\Omega}$:
\begin{equation}
\vec{\mathcal{J}}~=\vec{\mathcal{L}}~+~\vec{\mathcal{S}}~\sim~\vec{\Omega}/|\vec{\Omega}|
\end{equation}
where $\vec{\mathcal{L}}$ and $\vec{\mathcal{S}}$ are the total orbital 
and spin angular momenta, respectively. In case of the pion superfluidity
(and for a certain range of $|\vec{\Omega}|$) the alignment of $\vec{\mathcal{L}}$ to $\vec{\Omega}$
was demonstrated long time ago, see textbooks \cite{Landau:1959} 
and a brief discussion above.
The total-spin part in this model is represented by the spin of the cores of the vortices,
see for discussion above. Moreover, the spin angular momentum represents a quantum
correction, as follows from the quantization condition (\ref{quantization}).
What is revealed by the consideration of the chiral vortical effect,
see \cite{Teryaev:2017,Kirilin:2012} and discussion above,
is that the spin part $\vec{\mathcal{S}}$ is dual to the anomalous triangle graph.
It is this demonstration of the duality between the triangle graph and
spin of the vortices that makes definite the notion of the ``anomalous spin''
which we are introducing here.
 
\subsection{Field theoretic duality between IR and UV degrees of freedom}

Field theory allows to reduce the chiral vortical effect to the matrix element
(\ref{extraterm}). One can estimate this matrix element within effective theory
determining couplings of pions to heavy degrees of freedom, say, hyperons .
\cite{Teryaev:2017}. Actually, this matrix element is saturated by contribution of
heavy particles and, as a result, the spin of vortices can be related to the 
 spin of hyperons. Note that hyperons are not originally
included into the effective theory (\ref{lchirallagrangian}).
However, the UV completion of the effective theory turns necessary
to evaluate the matrix element (\ref{extraterm}). As a result, the chiral vortical
effect, evaluated originally for massless fermions, gets related to average
value of spin of
heavy hyperons \cite{Teryaev:2017}\footnote{ The dissipation 
makes the link with other QCD approaches to polarization where absorptive phases are necessary.}.

The construction might seem too far fetching to be true. It is amusing therefore
to note that a similar connection between UV and IR degrees of freedom 
apparently
holds in case of superfluid
$^3 He-A$\cite{Kita:1998,Volovik:2000}. The effective theory contains massless fermions
which are collective excitations while the UV degrees of freedom are provided by the
atoms of $^3He$. One can demonstrate explicitly that the density 
of angular momentum can be expressed both in terms of the IR and UV 
 (collective or corresponding to atoms)  degrees of freedom.
The system is nonrelativistic and all the calculations are explicit and under control.
 
\subsection{Acknowlegments} 

 We are thankful to P.G. Mitkin, A.V. Sadofyev, A. V. Vasiliev for useful discussions.

The work was supported by RFBR grant 17-02-01108.

\end{document}